\documentclass[reqno]{article}
\usepackage{amssymb,amsmath,epsfig,cite}

\begin{document}
\begin{center}{\Large \bf The Four Dimensional Dirac Equation in Five Dimensions}
\vskip.25in
Romulus Breban\\
{\it Institut Pasteur, Paris, France} 
\end{center}

\begin{abstract}
The Dirac equation may be thought as originating from a theory of five-dimensional (5D) space-time. We define a special 5D Clifford algebra and introduce a spin-1/2 constraint equation to describe null propagation in a 5D space-time manifold.  We explain how the 5D null formalism breaks down to four dimensions to recover two single-particle theories. Namely, we obtain Dirac's relativistic quantum mechanics and a formulation of statistical mechanics.  Exploring the non-relativistic limit in five and four dimensions, we identify a new spin-electric interaction with possible applications to magnetic resonance spectroscopy (within quantum mechanics) and superconductivity (within statistical mechanics).
\end{abstract}

\section{Introduction}\label{sec:intro}

The Dirac equation \cite{Dirac:1928vg,Dirac:1928wh} established itself as a fundamental tool in four dimensional (4D) physics. With increasing popularity of 5D physics, many authors generalized it from four to five dimensions \cite{Schouten:1932a,Pauli:1933ww,Dirac:1935wi,Lubanski:1942wu,Lubanski:1942ei,Bhabha:1945vt,Gursey:1963ve,Kocinski:1999tv,Zhang:2000bn,Redington:2007wy}. A different approach was that of Barut \cite{Barut:1964cb} who recognized that the Dirac matrices $\alpha^j$ ($j,k, ...=1,2,3$) and $\beta$ close under commutation to provide a representation of so(1,4), thus establishing an intrinsic 5D symmetry of the Dirac equation. Later, Bracken and Cohen \cite{Bracken:1969:CTD} introduced an algebraic formalism, for the Dirac equation, based on so(1,4). In this paper, we present a curved geometry for 5D space-times compatible to the Dirac equation, where the electromagnetic field enters according to the minimal coupling recipe. Furthermore, we reconsider the non-relativistic limit of the Dirac equation using 5D-based approximations.

We proceed by defining a special 5D Clifford algebra, using the O(1,4) symmetry, and propose a geometric constraint for 5D null propagation of spin-1/2 particles. We then use a recent interpretation of the 5D geometry \cite{Breban:2015bf,Breban:2005wf,Breban:2016ci} where a 5D null constraint yields a 4D on-shell constraint.  For example, the norm of the 5-momentum of a 5D photon in flat space-time $\eta_{AB}p^A p^B=0$ [$A, B = 0,1,2,3,5$, and $\eta_{AB}=diag(-1,1,1,1,1)$] yields the 4D on-shell constraint for a particle of mass $m$, $\eta_{\alpha\beta}p^\alpha p^\beta=-m^2c^2$ ($\alpha,\beta = 0,1,2,3$), where we used $p^5\equiv mc$. The mass $m$ is thought here as being {\it passive} \cite{Jammer:1997vj,Jammer:2009uk}, a dynamical variable which is not a source of gravitational field.\footnote{Mass which is a source of gravitational field is called {\it active} \cite{Jammer:1997vj,Jammer:2009uk}. The dichotomy between active and passive mass is not frame invariant. Here we consider that an observer may use any frame, except the comoving frame of the particle.} With this interpretation, the inverse Compton wavelength $\lambda^{-1}=mc/\hbar$ and the fifth coordinate $x^5$ are conjugated by Fourier transform. Furthermore, it is postulated that, although the geometry is 5D, the perception of an observer remains 4D. However, the 4D observer has two major tools for understanding the 5D space-time. If the 5D geometry is independent of the fifth coordinate, then the 5D physics can be interpreted as 4D quantum mechanics. If the 5D geometry is independent of time (i.e., the zeroth coordinate), then the 5D physics can be interpreted as 4D statistical mechanics \cite{Breban:2015bf,Breban:2005wf,Breban:2016ci}. Both mechanics and statistics pictures apply when the geometry is independent on both time and fifth coordinate.  In what follows, we consider how a 5D equation delivers the Dirac equation in 4D quantum mechanics and applications to 4D statistical mechanics.

We are particularly interested in a certain class of curved 5D space-times with the 4D interpretation of a Minkovski space with superimposed electromagnetic field. These space-times should be regarded as local approximations of more realistic 5D geometries, including non-trivial gravitational fields, whose metrics satisfy suitable field equations \cite{Breban:2005wf, Wesson:1999ta}. To describe spin-1/2 propagation in these space-times, one would naturally use the Dirac equation, where the electromagnetic field appears according to the minimal coupling recipe \cite{Dirac:1928vg}. Hence, our discussion of the Dirac equation within Kaluza-Klein theory puts into perspective generalizations of the Dirac equation for curved manifolds \cite{Gilbert:1991uq,Wald:2010un}.

The structure of this paper is as follows. In Sec.~\ref{sec:Cliff} we introduce a special Clifford algebra for a 5D flat space-time. In Sec.~\ref{sec:funf} we discuss a class of 5D curved space-times and postulate a spin description of 5D null propagation. In Secs.~\ref{sec:Dirac} and \ref{sec:5DDiracSM}, we discuss the relation with the 4D formalism and work out examples of physical systems. In Sec.~\ref{sec:conclusions} we conclude our work.

\section{A special 5D Clifford algebra}\label{sec:Cliff}

We use bold face to indicate that some indices of the corresponding mathematical objects are suppressed. Consider the vector space 
\begin{eqnarray}
\mathbb{H}^{1,1}=\left\{\boldsymbol\Psi=\left(\begin{array}{c}\Psi_1\\ \Psi_2\end{array}\right)\bigg|\:\Psi_{1,2}\in\mathbb{H}\right\} 
\end{eqnarray}
consisting of quaternion vectors with two components, henceforth called {\it biquaternions}.  The scalar product of two biquaternions $\boldsymbol\Psi,\boldsymbol\Phi\in\mathbb{H}^{1,1}$ is defined by
\begin{eqnarray}
\boldsymbol\Psi\cdot\boldsymbol\Phi\equiv\Psi^{\dagger a}\Gamma_{ab}\Phi^b.
\end{eqnarray}
The symbol $\dagger$ stands for hermitian conjugation and $\Gamma_{ab}$ ($a,b,...=1,2$ or $1,2,3,4$, as needed)\footnote{The indices $a,b,...$ run over the dimension of the $\boldsymbol\Gamma$ matrix. In turn, this depends on the algebraic representation of the quaternion units. Hamilton's representation requires $a,b,...=1,2$. The representation based on the Pauli matrices requires $a,b,...=1,2,3,4$.} for a hermitian matrix with quaternion entries and signature $(+,-)$. The isometry group of $\mathbb{H}^{1,1}$ is Sp(1,1), consisting of all transformations which preserve the norm and products of biquaternions. 

Using that Sp(1,1) is the minimal compact cover of O(1,4) (i.e., O(1,4)$\sim$Sp(1,1)/Z$_2$), we introduce a special Clifford algebra, ${\mathcal Cl^s}_{1,4}$, such that its basis elements $\Gamma^{Aa}{}_b$ satisfy the following relations 
\begin{eqnarray}
\label{eq:Gdef1}
\boldsymbol\Gamma^A \boldsymbol\Gamma^B-\boldsymbol\Gamma^B \boldsymbol\Gamma^A &=&-2 i\boldsymbol\sigma^{AB},\\
\label{eq:Gdef2}
\Gamma^{Aa}{}_b\Gamma^{Bb}{}_c+\Gamma^{Ba}{}_b\Gamma^{Ab}{}_c&=&2\eta^{AB}1^a{}_c,\\
\label{eq:Gdef3}
\boldsymbol\Gamma^0\boldsymbol\Gamma^1\boldsymbol\Gamma^2\boldsymbol\Gamma^3\boldsymbol\Gamma^5&=&i\boldsymbol1,
\end{eqnarray}
where $i$ is the complex unit, $\boldsymbol\sigma_{AB}$ is the so(1,4) generator of the rotation or boost in the plane $(A,B)$ and $\boldsymbol 1$ is the unit matrix. We further request 
\begin{eqnarray}
\boldsymbol\Gamma_0^\dagger=-\boldsymbol\Gamma_0, \quad \boldsymbol\Gamma_j^\dagger=\boldsymbol\Gamma_j, \quad \boldsymbol\Gamma_5^\dagger=\boldsymbol\Gamma_5.  
\label{eq:hermitian}
\end{eqnarray}
We note that the $\Gamma$ matrices are not determined by Eqs.~\eqref{eq:Gdef1}-\eqref{eq:hermitian}, since a conjugacy (or similarity) transformation with a unitary matrix $\boldsymbol U$ (i.e., $\boldsymbol U^\dagger=\boldsymbol U^{-1}$), $\boldsymbol\Gamma^A\rightarrow \boldsymbol U^{-1}\boldsymbol\Gamma^A\boldsymbol U$, yields another set of matrices satisfying Eqs.~\eqref{eq:Gdef1}-\eqref{eq:hermitian}. Two representations of the $\Gamma$ matrices are provided in Appendix \ref{Appendix:A}.

The cover relationship between Sp(1,1) and O(1,4) can be made explicit using the $\Gamma$ matrices. For every $\boldsymbol S\in$~Sp(1,1), there exists $\boldsymbol\Lambda\in$~SO(1,4) such that 
\begin{eqnarray}
\label{eq:cover}
\boldsymbol S^{-1}\boldsymbol\Gamma_A\boldsymbol S=\Lambda_A{}^B\boldsymbol\Gamma_B.
\end{eqnarray}
In contrast, given $\boldsymbol\Lambda\in$~O(1,4), $\boldsymbol S\in$~Sp(1,1) is determined up to a sign. Still, SO(1,4) and Sp(1,1) share the same Lie algebra. An infinitesimal Sp(1,1) transformation, $\boldsymbol S=1-i\boldsymbol\sigma_{AB}w^{AB}/4$, where $w_{AB}$ is an infinitesimal antisymmetric form, corresponds to an infinitesimal SO(1,4) transformation, written as $\Lambda_A{}^B=\eta_A{}^B+w_A{}^B$. The condition that an infinitesimal Sp(1,1) transformation leaves $\boldsymbol\Psi\cdot\boldsymbol\Phi$ invariant yields
\begin{eqnarray}
\label{eq:G12}
\boldsymbol\Gamma\boldsymbol\sigma^{\alpha\beta}-\boldsymbol\sigma^{\alpha\beta}\boldsymbol\Gamma&=&\boldsymbol0,\\
\label{eq:G22}
\boldsymbol\Gamma\boldsymbol\sigma^{\alpha 0}+\boldsymbol\sigma^{\alpha 0}\boldsymbol\Gamma&=&\boldsymbol0,
\end{eqnarray}
where $\alpha,\beta=1,2,3,5$.
$\boldsymbol\Gamma$ is a solution of Eqs.~\eqref{eq:G12} and \eqref{eq:G22} only if it is a linear combination (with complex coefficients) of $\boldsymbol\Gamma^0$ and $\boldsymbol\Gamma^1\boldsymbol\Gamma^2\boldsymbol\Gamma^3\boldsymbol\Gamma^5$.\footnote{In this context, ${\mathcal Cl^s}_{1,4}$ is naturally generated by the hermitian biquaternion metric $\boldsymbol\Gamma$ and the Sp(1,1) generators, $-i\boldsymbol\sigma^{AB}$.} This concludes the construction of ${\mathcal Cl^s}_{1,4}$. By convention, we use $\boldsymbol\Gamma$ and its inverse to lower and raise the $a, b, ...$ indices. We also note the identity $\boldsymbol\Gamma\boldsymbol\Gamma^\dagger_A=-\boldsymbol\Gamma_A\boldsymbol\Gamma$ and that $(\boldsymbol\Gamma)^2\propto \boldsymbol 1$.

\section{On 5D space-time and 5D null propagation}\label{sec:funf}

We discuss a particular class of sufficiently smooth 5D Lorentzian manifolds, equipped with a space-like Killing field, subject to the principle of general covariance. We denote such a manifold by $\mathcal{M}$ and use indices $M, N,\dots=0,1,2,3,5$ for the global 5D geometry, while $A, B,\dots=0,1,2,3,5$ stand for the local 5D geometry (i.e., tangent spaces). Furthermore, we use the indices $\mu,\nu,\dots=0,1,2,3$ for foliated curved 4D metrics and the indices $\alpha,\beta,\dots=0,1,2,3$ for foliated flat 4D metrics.

We denote the 5D metric by $h_{MN}$, choose the fifth coordinate $x^5$ along the Killing field and foliate the 5D space-time along $x^5$. We denote the foliated 4D metric by $g_{\mu\nu}$ and the shift of the foliation by $N_\mu$. Since we are only interested in 5D null propagation, we set the lapse of the foliation to 1 by a conformal transformation. The transformed 5D metric is denoted by $\tilde h_{MN}$
\begin{eqnarray}
\label{eq:metric}
\tilde h^{MN}=\left(\begin{array}{cc} \tilde g^{\mu \nu} & N^\mu \\
                                  N^\nu  &  N^\rho N_\rho +1
\end{array} \right),\quad 
\tilde h_{MN}=\left(\begin{array}{cc}
\tilde g_{\mu \nu}+N_\mu N_\nu & -N_\mu \\
             -N_\nu  & 1
\end{array}\right),
\end{eqnarray}
where $\tilde g_{\mu\nu}$ stands for the transformed 4D metric.

To describe particle propagation, we choose a local 5D Lorentzian frame $e^A{}_M$, called {\it f\"unfbein}, at each point of the 5D manifold
\begin{eqnarray}
\tilde h_{MN}=e^A{}_Me^B{}_N\,\eta_{AB},\quad\eta_{AB}=e^M{}_Ae^N{}_B\,\tilde h_{MN}.
\end{eqnarray}
We further provide ${\mathcal Cl^s}_{1,4}$ structure to the tangent space $\mathcal{T}_p$ at each point $p$ of the manifold. That is, the f\"unfbein components of a tangent vector $x^A$ at $p$ yield an element $\boldsymbol x=x^A\boldsymbol\Gamma_A$ of a locally defined ${\mathcal Cl^s}_{1,4}$ algebra.\footnote{The square norm of a Clifford element $\boldsymbol x$ is given by the product $\boldsymbol x\, \boldsymbol x^\dagger=\boldsymbol x^\dagger \boldsymbol x= \sum_A (x^A)^2$, inducing a 5D Euclidian metric.} An element of $\mathcal{T}_p$ transforms with SO(1,4) (i.e., $x^B\rightarrow\Lambda_A^B x^A$), while an element of ${\mathcal Cl^s}_{1,4}$ transforms with Sp(1,1)  (i.e., $\boldsymbol x\rightarrow\boldsymbol S^{-1}\boldsymbol x \boldsymbol S$). The biquaternion metric $\boldsymbol\Gamma$ may change over the 5D space-time and is considered on equal footing with the  f\"unfbein metric, $\eta_{AB}$.

In particular, we wish to describe the experience of {\it uncharged} local observers, who perceive geometrically the first four space-time dimensions, but not the fifth.  We choose the following non-orthogonal f\"unfbein to reflect this
\begin{eqnarray}
\label{eq:funfb}
e^A{}_M=\left(\begin{array}{cc}f^\alpha{}_\mu&\boldsymbol 0\\-N_\mu&1\end{array}\right),\quad
e^M{}_A=\left(\begin{array}{cc}f^\mu{}_\alpha&\boldsymbol 0 \\  N_\mu f^\mu{}_\alpha&1\end{array}\right),
\end{eqnarray}
where $\boldsymbol 0$ denotes here a $4\times 1$ matrix with null entries, and $f^\alpha{}_\mu$ is a local Lorentzian frame (vierbein) at each point of the 4D space-time with the metric $\tilde g_{\mu\nu}$. We may further ask that the 5D manifold be interpreted as a conformally flat 4D space-time with superimposed electromagnetic fields.  In this case, the foliated 4D metric is $\tilde g_{\mu\nu}=\eta_{\mu\nu}$ and we choose $f^\alpha{}_\mu$ to be the Kronecker-delta tensor; i.e., $f^\alpha{}_\mu=1^\alpha{}_\mu$. Furthermore, in cgs units, we have $N_\mu=-qA_\mu/c^2$, where $q$ is specific charge, $A_\mu$ is the electromagnetic potential, and $c$ is the speed of light in vacuum.

We describe 5D null propagation by assigning two 5D null vectors (or their representations), one ingoing and one outgoing, at each point of the 5D manifold. Point-to-point infinitesimal propagation may be described by infinitesimal translation followed by suitable, smooth isometric transformations of the null vectors. Hence, we impose smoothness conditions. We would like each null vector be smoothly cast into any other null vector, by local isometry transformations.\footnote{In the 4D picture, this would imply smooth transformations between particle and antiparticle states.} This is not possible since O(1,4) is not a connected Lie group. Hence, we represent each pair of 5D null vectors as a biquaternion,\footnote{A pair of 5D null vectors has the same number of real degrees of freedom as a biquaternion.} and describe 5D null propagation using a map $\boldsymbol\Psi:\mathcal{M}\rightarrow\mathbb{H}^{1,1}$. The biquaternion $\boldsymbol\Psi(x^M)$ at point $p$, with coordinates $x^M$, transforms with elements of Sp(1,1) [i.e., $\boldsymbol \Psi(x^M)\rightarrow\boldsymbol S\boldsymbol \Psi(x^M)$], which is simply connected and compact.

Next, we require the map $\boldsymbol\Psi(x^A)$ be Clifford differentiable \cite{Gilbert:1991uq} (i.e., differentiable with respect to $\boldsymbol x=x^A\boldsymbol\Gamma_A$) in every tangent space endowed with ${\mathcal Cl^s}_{1,4}$ structure.  This requirement constrains $\boldsymbol\Psi(x^A)$ to satisfy the following differential equation
\begin{eqnarray}
\label{eq:dirac0}
\Gamma^{Aa}{}_b\partial_A\Psi^b(x^A)=0^a.
\end{eqnarray}
As it turns out, the constraint \eqref{eq:dirac0} is already sufficient for describing 5D particle propagation. We are particularly interested how $\boldsymbol\Psi(x^A)$, solution of \eqref{eq:dirac0}, behaves on the local null cone ${\mathcal C_p}=\{x^A\in \mathcal{T}_p; \boldsymbol x=x^A\boldsymbol\Gamma_A \in {\mathcal Cl^s}_{1,4}|\,\boldsymbol x^2=0\}$. We may write the derivative in the local tangent space as $\partial_A=e_A{}^M\partial_M$ and cast Eq.~\eqref{eq:dirac0} over an open neighborhood of the 5D manifold
\begin{eqnarray}
\label{eq:dirac}
(\Gamma^{Aa}{}_be_A{}^M)\partial_M\Psi^b(x^M)=0^a.
\end{eqnarray}
This completes our spin description of 5D null propagation, as an alternative to the null path-integral formulation \cite{Breban:2005wf,Breban:2015bf,Breban:2016ci}. Dirac operators over curved manifolds are commonly introduced using standard Clifford algebras and spin connections \cite{Gilbert:1991uq}. These mathematical constructs do not appear in Eq.~\eqref{eq:dirac}.

In the applications we present below, we break wavefunction covariance and use metrics $\tilde h_{MN}$ that are independent of both $x^5$ and $x^0$.  Hence, we derive 5D equations for quaternionic wavefunctions, starting from Eq.~\eqref{eq:dirac} rewritten as
\begin{eqnarray}
\label{eq:Dirac_SM}
\boldsymbol\Gamma^5\partial_5\boldsymbol\Psi+\boldsymbol\Gamma^\alpha\partial_\alpha\boldsymbol\Psi+\boldsymbol\Gamma^\alpha N_\alpha\partial_5\boldsymbol\Psi=0, 
\end{eqnarray}
where $N_\alpha\equiv N_\mu f^\mu{}_\alpha$. In the primed representation for the $\Gamma$ matrices (Appendix \ref{Appendix:A}), we obtain
\begin{eqnarray}
\label{eq:Psi1}
(\partial_5+\partial_0+N_0\partial_5)\boldsymbol\Psi_2-\boldsymbol\sigma^j(\partial_j+N_j\partial_5)\boldsymbol\Psi_1=0,\\
\label{eq:Psi2}
(\partial_5-\partial_0-N_0\partial_5)\boldsymbol\Psi_1+\boldsymbol\sigma^j(\partial_j+N_j\partial_5)\boldsymbol\Psi_2=0,
\end{eqnarray}
where $\boldsymbol\sigma^j$ are the Pauli matrices and $\boldsymbol\Psi_{1,2}$ are the quaternionic components of $\boldsymbol\Psi$. We impose\footnote{In this case, the Lorentz and Coulomb gauges are equivalent.} $\partial_{0}N_\alpha=0$ and $\partial_{5}N_\alpha=0$, and apply the operator $\boldsymbol\sigma^j(\partial_j+N_j\partial_5)$ on the left of Eqs.~\eqref{eq:Psi1} and \eqref{eq:Psi2}, mindful of the commutator
\begin{eqnarray}
[\boldsymbol\sigma^j(\partial_j+N_j\partial_5),(\partial_5\pm\partial_0\pm N_0\partial_5)]=\pm\boldsymbol\sigma^j(\partial_j N_0)\partial_5.
\end{eqnarray} 
Introducing the wavefunctions $\boldsymbol\Psi_{\pm}=\boldsymbol\Psi_1\pm\boldsymbol\Psi_2$, Eqs.~\eqref{eq:Psi1} and \eqref{eq:Psi2} yield
\begin{eqnarray}
[\partial_5^2-(\partial_0+N_0\partial_5)^2]\boldsymbol\Psi_{\pm}+[\boldsymbol\sigma^j(\partial_j+N_j\partial_5)]^2\boldsymbol\Psi_{\pm}\mp\boldsymbol\sigma^j(\partial_j N_0)\partial_5\boldsymbol\Psi_{\pm}=0.
\label{eq:diracSpin}
\end{eqnarray}
It is interesting to contrast Eq.~\eqref{eq:diracSpin} to the corresponding 5D Klein-Gordon (KG) equation \cite{Breban:2015bf}
\begin{eqnarray}
\tilde\bigtriangledown_M\tilde\bigtriangledown^M\Psi=0\Leftrightarrow[\partial_5^2-(\partial_0+N_0\partial_5)^2]\Psi+[(\partial^j+N^j\partial^5)(\partial_j+N_j\partial_5)]\Psi=0;
\label{eq:5DKG}
\end{eqnarray}
$\tilde\bigtriangledown_M$ is the covariant derivative of the metric $\tilde h_{MN}$ \eqref{eq:metric} with $\tilde g_{\mu\nu}=\eta_{\mu\nu}$, and we imposed only $\partial_5N_\mu=0$.  The operator $\partial_5^2-(\partial_0+N_0\partial_5)^2$ is key for the non-relativistic limit and common to both Eqs.~\eqref{eq:diracSpin} and \eqref{eq:5DKG}, virtue to the representation of the $\Gamma$ matrices.  

The non-relativistic limit is thought as the limit of vanishing kinetic energy or low temperatures. For a particle propagating with null 5-momentum $p_A$ (i.e., a 5D photon) in flat 5D space-time, we impose $|p_j|\ll |p_0|$ and $|p_j|\ll |p_5|$, which yield $p_0\approx \pm p_5$. Hence, a particle starting at the origin of the coordinate frame remains localized around $x^0\approx \pm x^5$. The propagation problem reduces to the 5D KG equation, which, in light-cone coordinates, 
\begin{eqnarray}
y^0=x^5-x^0,\quad y^j=x^j,\quad y^5=(x^0+x^5)/2,
\label{eq:changeQM}
\end{eqnarray}
leads to the Schr\"odinger and the Fokker-Planck (FP) equations \cite{Breban:2015bf}.  In the case of weak fields, Eq.~\eqref{eq:changeQM} still provides approximate light-cone coordinates where
\begin{eqnarray}
\partial_5^2-(\partial_0+N_0\partial_5)^2\approx 2\frac{\partial^2}{\partial y^0\partial y^5}-2N_0\partial_0\partial_5.
\label{eq:5Dnr}
\end{eqnarray}
The operators $\partial_{0}$ and $\partial_{5}$ in Eqs.~\eqref{eq:diracSpin} and \eqref{eq:5Dnr} may be explicitly written as 
\begin{eqnarray}
\partial_0=-\frac{\partial}{\partial y^0}+\frac{1}{2}\frac{\partial}{\partial y^5},\quad \partial_5=\frac{\partial}{\partial y^0}+\frac{1}{2}\frac{\partial}{\partial y^5},
\end{eqnarray}
subject to further approximations, depending on the application.

\section{The Dirac equation}\label{sec:Dirac}

We now explain how Eq.~\eqref{eq:dirac} relates to the Dirac equation. The biquaternion $\boldsymbol\Psi$ can be cast into a bispinor pair; see Appendix \ref{Appendix:B}.  Hence, solving Eq.~\eqref{eq:dirac} reduces to solving 
\begin{eqnarray}
\label{eq:dirac2}
e^M{}_A\Gamma^{Aa}{}_b\partial_M\psi^b(x^M)=0^a,
\end{eqnarray}
where $\boldsymbol\psi$ is a bispinor. Using Eq.~\eqref{eq:funfb}, left-multiplying by $i \hbar\boldsymbol\Gamma^5$, and performing a Fourier transform with respect to $x^5$ (n.b., $x^5$ is Fourier conjugated to $\lambda^{-1}=mc/\hbar$), we obtain a Dirac-type equation describing propagation in a curved 4D space-time with superimposed electromagnetic fields
\begin{eqnarray}
\label{eq:Dirac}
\boldsymbol\gamma^\alpha(i\hbar f^\mu{}_\alpha\partial_\mu+qmA_\alpha/c)\hat{\boldsymbol\psi}-mc\hat{\boldsymbol\psi}=\boldsymbol 0,
\end{eqnarray}
where we used hat $\hat{}$ to denote the Fourier transforms of 5D wavefunctions with respect to $x^5$.  The matrices $\boldsymbol\gamma^\alpha\equiv\boldsymbol\Gamma^5\boldsymbol\Gamma^{\alpha}$ form the basis for a foliated 4D Clifford algebra where
\begin{eqnarray}
\label{eq:gamma}
\gamma^{\alpha a}{}_b\gamma^{\beta b}{}_c+\gamma^{\beta a}{}_b\gamma^{\alpha b}{}_c=-2\eta^{\alpha\beta}1^a{}_c.
\end{eqnarray}
 In this case, rather than requiring Eq.~\eqref{eq:Dirac} be defined on the 5D null structure, we impose an {\it on-shell constraint} for the propagation in the 4D space-time resulting from foliating $x^5$. In general, this constraint depends on the electromagnetic field.\footnote{In each tangent space of the 4D manifold with electromagnetic fields, we have $\eta_{\alpha\beta}p^\alpha p^\beta=-m^2c^2$. However, this result cannot be extended over the manifold without accounting for the electromagnetic field.} If the foliated 4D metric is flat (i.e.,
$\tilde g_{\mu\nu}=\eta_{\mu\nu}$) and we choose $f^\alpha{}_\mu=1^\alpha{}_\mu$, then Eq.~\eqref{eq:Dirac} becomes the Dirac equation \cite{Dirac:1928vg} where the electromagnetic field enters according to the minimal coupling recipe
\begin{eqnarray}
\label{eq:Dirac_}
\boldsymbol\gamma^\alpha(i\hbar\partial_\alpha+qmA_\alpha/c)\hat{\boldsymbol\psi}-mc\hat{\boldsymbol\psi}=\boldsymbol 0.
\end{eqnarray}

A gauge transformation of the electromagnetic field may be introduced in Eq.~\eqref{eq:dirac} through a 5D transformation of coordinates: $x^\mu\rightarrow x^\mu,\, x^5\rightarrow x^5+\mathcal{N}(x^\mu)$. If the transformation is passive, it applies to the 5D manifold coordinates, resulting in a transformation for the shift of the foliation $N_\mu\rightarrow N_\mu+\partial_\mu\mathcal{N}$, which is proportional to the electromagnetic field and satisfies the Maxwell equations \cite{Breban:2005wf}. If the transformation is active, then one changes 
\begin{eqnarray}
\label{eq:EMgauge}
\boldsymbol\Psi(x^\mu,x^5)\rightarrow\boldsymbol\Psi(x^\mu,x^5+\mathcal{N}(x^\mu)). 
\end{eqnarray}
A Fourier transform of Eq.~\eqref{eq:EMgauge} with respect to $x^5$ yields 
\begin{eqnarray}
\hat{\boldsymbol\Psi}\rightarrow\hat{\boldsymbol\Psi}\exp[i\mathcal{N}(x^\mu)\lambda^{-1}]\equiv\hat{\boldsymbol\Psi}\exp\left[-i\frac{qm}{\hbar c}\mathcal{A}(x^\mu)\right],
\end{eqnarray}
where $\mathcal{A}(x^\mu)$ is the gauge function of the electromagnetic field. Hence, as expected, the electromagnetic field can be interpreted as a U(1)-gauge field for Eq.~\eqref{eq:Dirac} and the Dirac equation \eqref{eq:Dirac_}, but not the 5D Dirac equation \eqref{eq:dirac}.  This formalism may be developed further using the 4D Clifford algebra of the $\gamma$ matrices \eqref{eq:gamma} and the picture of a 4D space-time with superimposed electromagnetic field \cite{Thaller:2013dl}.

Derivations of the Dirac equation from first physics principles, in both flat and curved 4D space-times were previously discussed \cite{Wald:2010un,Srinivasan:1996uw,Noyes:uk,Ng:2003p1101,Deriglazov:2014cp}. Still, the Dirac equation has been criticized for the Klein paradox \cite{Alhaidari:2009kp,Calogeracos:1999ew,MIYAZAKI:tz,0143-0807-36-5-055015} and causality violation \cite{Barat:2003gp,Castrigiano:2015ck,1985PhRvL..54.2395H,Rosenstein:1987uj} in quantum mechanics. The Klein paradox may be solved by imposing smooth interaction potentials (i.e., smooth space-time manifold) for the Dirac equation \cite{Calogeracos:1999ew}. Our 5D approach revisits the geometric interpretation of the Dirac equation, bringing additional physics into play. First, the particle (passive) mass appears through Fourier transform with respect to $x^5$. Second, electromagnetic (minimal) coupling enters naturally the Dirac equation, virtue to the 5D geometry. Furthermore, arguments of causality violation brought against the Dirac equation assume that propagation is fundamentally 4D, locally confined to the interior of 4D null cones. They do not hold when considering that the Dirac equation originates from a description of 5D null propagation.

\subsection{On the non-relativistic limit of the Dirac equation}

It is well-known that the non-relativistic limit of the Dirac equation results as the Pauli equation, when separating {\it small} and {\it large} components of the bispinor wavefunction $\hat{\boldsymbol\psi}$ \cite{Schwabl:2005xy,Hecht:2000ul}. Here, we do not employ this technique. Rather, we use 5D geometric-based approximations to obtain equations for the non-relativistic propagation of a quantum particle (i.e., $p^5>0$) in time-independent, weak fields. We consider that $y^5\approx x^5$ and $p_0\approx p_5$ (i.e., $\partial_0\boldsymbol\Psi_{\pm}\approx\partial_5\boldsymbol\Psi_{\pm}$) in the terms containing fields in Eqs.~\eqref{eq:diracSpin} and \eqref{eq:5Dnr} to obtain
\begin{eqnarray}
2\left(\frac{\partial}{\partial y^0}\partial_5-N_0\partial_5^2\right)\boldsymbol\Psi_{\pm}+[\boldsymbol\sigma^j(\partial_j+N_j\partial_5)]^2\boldsymbol\Psi_{\pm}\mp\boldsymbol\sigma^j(\partial_j N_0)\partial_5\boldsymbol\Psi_{\pm}\approx0.
\label{eq:diracSpin_}
\end{eqnarray}
We perform a Fourier transform of Eq.~\eqref{eq:diracSpin_} with respect to $x^5$ and use $y^0\equiv c\tau$ as time coordinate, conjugated with the mechanical energy, only; i.e., not including $mc^2$. Furthermore, we cast the wavefunction $\hat{\boldsymbol\Psi}_{\pm}$ into a spinor pair (Appendix \ref{Appendix:B})  for a non-relativistic picture of spin 1/2 quantum mechanics, based on the $L_2({\mathbb R}^3)\times L_2({\mathbb R}^3)$ Hilbert space equipped with the canonical dot product. In typical notation of 3D physics, this yields
\begin{eqnarray}
-\frac{\hbar}{i}\frac{\partial\hat{\boldsymbol\psi}_{\pm}}{\partial\tau}\approx
\frac{1}{2m}\left(\frac{\hbar}{i}\nabla-\frac{e}{c}\vec{A}\right)^2\hat{\boldsymbol\psi}_{\pm}+eA_0\hat{\boldsymbol\psi}_{\pm}-\mu\vec{\boldsymbol\sigma}\vec{B}\hat{\boldsymbol\psi}_{\pm}
\mp i\mu\vec{\boldsymbol\sigma}\vec{E}\hat{\boldsymbol\psi}_{\pm},
\label{eq:Pauli+}
\end{eqnarray}
where $\hat{\boldsymbol\psi}_\pm$ is the spinor wavefunction, $\vec{B}=\nabla\times\vec{A}$ and $\vec{E}=-\nabla{A_0}$ are time-independent magnetic and electric fields, and $e\equiv qm$ and $\mu\equiv q\hbar /(2c)$ are, respectively, the particle's charge and magneton. Equation \eqref{eq:Pauli+} represents the Pauli equation for time-independent settings, except for the term $\mp i\mu\vec{\boldsymbol\sigma}\vec{E}\hat{\boldsymbol\psi}_{\pm}$, describing an interaction between spin and time-independent electric field. An anti-hermitian spin-electric interaction term occurs next to the well-known Stern-Gerlach (hermitian) spin-magnetic term. Thus, the total spin-interaction term is given by $\mu\vec{\boldsymbol\sigma}(\vec{B}\pm i\vec{E})\hat{\boldsymbol\psi}_{\pm}$, where $\vec{B}\pm i\vec{E}$ is reminiscent of Riemann-Silberstein fields \cite{Silberstein:1907vk,BialynickiBirula:2013gs} and the photon wavefunction problem \cite{BialynickiBirula:1994tg,Popescu:2010ud}.

Spin-electric interactions are common in the phenomenology of condensed matter physics. They received much attention \cite{Rashba:2008ja,Li:2013cj}, particularly with the discovery of the electric-dipole spin resonance \cite{NadjPerge:2010kw,Yuan:2013bc}. Spin-orbit coupling and hyperfine interactions can mediate spin-electric interactions.  There are even cases where the Stern-Gerlach term contributes toward spin-electric interactions \cite{RamirezRuiz:2017jd}. In contrast to phenomenological contributions, $\mp i\mu\vec{\boldsymbol\sigma}\vec{E}\hat{\boldsymbol\psi}_{\pm}$ has fundamental status and is anti-hermitian, conventionally describing decoherence, decay and resonant scattering induced by electric fields on spin 1/2 particles. In fact, the process of breaking covariance of the Dirac equation \eqref{eq:Dirac_} down to Eq.~\eqref{eq:Pauli+} can be naturally explained in the language of non-hermitian quantum mechanics \cite{Rivas:1399189,Moiseyev:2011tx,Castagnino:ty}. 

The Dirac operator is essential self-adjoint \cite{Anonymous:-Q0trXeC,Arrizabalaga:2013fk,GEORGESCU:2006kz,Guneysu:2013ed,Kalf:2000kp}, leading to real energy spectra for many field configurations. Hence, the Dirac particle, described by the bispinor wavefunction $\hat{\boldsymbol\psi}$, represents a {\it closed system}. Due to covariance breaking, $\hat{\boldsymbol\psi}$ is rearranged into the spinor wavefunctions $\hat{\boldsymbol\psi}_\pm$.\footnote{For exact covariance breaking, without taking the non-relativistic limit, consider spinor solutions $\hat{\boldsymbol\psi}_\pm$ of the Fourier transform with respect to $x^5$ of Eq.~\eqref{eq:diracSpin}.} This results into two {\it open systems}, defined by non-hermitian hamiltonians on smaller Hilbert spaces. Each of these systems undergoes decoherence and consists of a Pauli particle in interaction with the {\it environment} through an anti-hermitian spin-electric term. However, the two open systems complete each other to form a larger closed system, which remains coherent.

From the perspective of the Pauli equation, Eq.~\eqref{eq:Pauli+} deserves attention in its own right. We discuss a simple application, neglecting the effect of electromagnetic fields on particle orbits, and analyzing only the spin degrees of freedom. Equation~\eqref{eq:Pauli+} yields
\begin{eqnarray}
-\frac{\hbar}{i}\frac{\partial\hat{\boldsymbol\psi}}{\partial\tau}\approx-\mu\vec{\boldsymbol\sigma}(\vec{B}
+ i\vec{E})\hat{\boldsymbol\psi}\equiv \boldsymbol H\hat{\boldsymbol\psi},
\label{eq:Pauli+spinonly}
\end{eqnarray}
where $\boldsymbol H$ is the hamiltonian operator acting on spinors $\hat{\boldsymbol\psi}$. We analyze Eq.~\eqref{eq:Pauli+spinonly} using the quantum theory of decay processes \cite{Fonda:1978dw}, but other approaches \cite{Zloshchastiev:2015ke} may apply, as well.  The probability\footnote{This quantity is not strictly a probability when hamiltonians are non-hermitian, since its value may exceed 1.} that state $\hat{\boldsymbol\psi}$, prepared at $\tau=0$, remains unaltered is given by
\begin{eqnarray}
P_{\hat\psi}(\tau)=\left|\hat{\boldsymbol\psi}^\dagger\! \exp(-i\boldsymbol H\tau/\hbar)\,\hat{\boldsymbol\psi}\right|^2.
\end{eqnarray}
Thus, within an ensemble of identical particles, the number $N(\tau)$ of particles in state $\hat{\boldsymbol\psi}$ at time $\tau$ is $N(\tau)=N(0)P_{\hat\psi}(\tau)$. If $P_{\hat\psi}(\tau)$ decreases with $\tau$, then we say that $\hat\psi$ undergoes spontaneous decay.  If $P_{\hat\psi}(\tau)$ increases with $\tau$, then we say that $\hat\psi$ gets populated.

In our case, we have
\begin{eqnarray}
\exp(-i\boldsymbol H\tau/\hbar)=\cos(\omega\tau)+\frac{i\vec{\boldsymbol\sigma}\vec{\omega}}{\omega}\sin(\omega\tau),
\end{eqnarray}
where $\vec{\omega}\equiv\mu(\vec{B}+ i\vec{E})/\hbar$ and $\omega\equiv\sqrt{(\vec{\omega})^2}$.  When $\vec{E}$ adds just a small perturbation to spin precession in the magnetic field $\vec{B}$, the first order expansion in $|\vec{E}|/|\vec{B}|$ yields $\omega\approx \mu(|\vec{B}|+ i|\vec{E}|\cos\theta)/\hbar$, where $\theta$ is the angle between $\vec{E}$ and $\vec{B}$.  The probabilities that spin up $(+)$ and spin down $(-)$ states, prepared with the magnetic field $\vec{B}$, remain unaltered are
\begin{eqnarray}
P_{\pm}(\tau)=\exp (\pm2\,\mathrm{Im}({\omega})\tau)=\exp (\pm2\mu|\vec{E}|\cos\theta\,\tau/\hbar).
\end{eqnarray}
Hence, spontaneous quantum transitions, due to the perturbing electric field $\vec{E}$, occur on the time scale 
\begin{eqnarray}
|2\,\mathrm{Im}({\omega})|^{-1}=\hbar/(2\mu|\vec{E}||\cos\theta|), 
\label{eq:broad}
\end{eqnarray}
which determines the coherence time and spectral width of the transition between spinor states prepared with the field $\vec{B}$. 

These transitions can be observed in resonant regime by augmenting the experimental setup corresponding to Eq.~\eqref{eq:Pauli+spinonly}  with the presence of a electromagnetic radiation field, for photons to be absorbed and reemited at the Larmor frequency, $\mu|\vec{B}|/\hbar$. This constitutes, in fact, the setup of electric-field nuclear magnetic resonance \cite{1986ZNatA..41..421R,1969MolPh..16..275H}. However, it appears that resonance broadening versus external electric field has not been addressed systematically. Nevertheless, particles in condensed state are often subject to {\it internal}, local electric fields.  It is well established that electric quadrupole interactions strongly broaden the nuclear magnetic resonance \cite{Freude:2006fg}, in qualitative aggreement with Eq.~\eqref{eq:broad}.

\section{On the 5D Dirac equation in statistical physics}\label{sec:5DDiracSM}

Quantum propagation has been previously invoked for a space-time picture of statistical physics.  The original formalism, proposed by Feynman \cite{Feynman:2010wb}, is based on path-integrals in a 4D space-time with Riemannian metric, where Kelvin temperature is used to construct the time coordinate; see Ref.~\cite{Wong:2014ea} for a recent review. In this case, the path integral between two arbitrary space-time events represents the propagator of a diffusion equation, which is interpreted as a quantum equation. Hence, the path integral is not indispensible and the formalism can be fully recovered from the diffusion equation alone.  

Here, general considerations and Eq.~\eqref{eq:dirac} yield a single-particle formalism for spin 1/2, in line with earlier work on 5D spinless propagation \cite{Breban:2005wf,Breban:2015bf}. We start with the observation that thermodynamic observables at thermal equilibrium do not change with time.  We assume that a particle achieves thermal equilibrium for a time-independent metric and impose $\tilde h^{MN}$ \eqref{eq:metric} be independent of both $x^5$ and $x^0$. In this case, the 5D metric can be foliated along $x^5$ or $x^0$, so physics can be interpreted from the perspective of mechanics or statistics, respectively.  The coordinate transformations that preserve this property are dipheomorphisms of 3D space (n.b.,~the local isometry group is O(3)). We further assume that a particle at thermal equilibrium is always found at rest by quantum measurement.

For connecting with previous results and mathematical convenience, we analyze the non-relativistic dynamics of a {\it statistical} particle in the presence of weak fields. Combining Eqs.~\eqref{eq:diracSpin} and \eqref{eq:5Dnr}, we obtain
\begin{eqnarray}
2\left(\frac{\partial^2}{\partial y^0\partial y^5}-N_0\partial_0\partial_5\right)\boldsymbol\Psi_{\pm}+[\boldsymbol\sigma^j(\partial_j+N_j\partial_5)]^2\boldsymbol\Psi_{\pm}\mp\boldsymbol\sigma^j(\partial_j N_0)\partial_5\boldsymbol\Psi_{\pm}\approx 0.
\label{eq:diracSpin__}
\end{eqnarray}
We introduce spinor rather than quaternionic wavefunctions, perform a Laplace transform with respect to $y^0$, and give $y^5\equiv cu$ the meaning of time.  Thus, searching for solutions in the form $\boldsymbol\psi_{\pm}(y^0,x^j,u)=\exp(-y^0/\Lambda)\boldsymbol\psi'_{\pm}(x^j,u)$ yields equations for $\boldsymbol\psi'_{\pm}(x^j,u)$, possibly with non-hermitian hamiltonians, where evolution is described by the dilatation group D(1), rather than U(1). Hence, these equations are interpreted as diffusion equations, where temperature enters indirectly through an empirical model for the diffusion constant. The fundamental solutions of the diffusion equations yield non-relativistic, single-particle propagators from $(\vec{x},0)$ to $(\vec{x},u)$, denoted $\boldsymbol\psi'_{\pm}(\vec{x},0; \vec{x},u)$. We use traces of these propagators over the space and spin degrees of freedom to define the Massieu function, $\Psi/k_B$, for a single particle at constant temperature
\begin{eqnarray}
\Psi/k_B=\ln \int d{\vec{x}}\,\,[\mathrm{Tr}\,\boldsymbol\psi'_+(\vec{x},0; \vec{x},u)+\mathrm{Tr}\,\boldsymbol\psi'_-(\vec{x},0; \vec{x},u)],
\label{eq:Massieu}
\end{eqnarray}
where the sum between $\mathrm{Tr}\,\boldsymbol\psi'_+(\vec{x},0; \vec{x},u)$ and $\mathrm{Tr}\,\boldsymbol\psi'_-(\vec{x},0; \vec{x},u)$ represents the trace over the spinor degrees of freedom. 

Thermodynamics require $\Psi/k_B$ be real-valued, which is not granted by Eq.~\eqref{eq:Massieu} and remains to be established for each particular application. When the Massieu function is well defined, thermodynamic results may be used to identify $u$ in Eq.~\eqref{eq:Massieu} as a time scale specific to the statistical particle.\footnote{The time-scale $u$ may be thought as the minimum amount of time for which a particle should be observed in order to be assigned thermodynamical observables. In Ref.~\cite{Breban:2005wf}, this is called a {\it quantum of physical time}.}  This is very different from current developments in non-hermitian quantum thermodynamics \cite{2010JPhA...43e5307J,Gardas:2016ia}, where the class of hamiltonians is carefully restricted such that the proposed formalism does not encounter difficulties.

Many-particle statistics may be approached through an ergodic principle.  We argue heuristically that single-particle propagation for the time interval $Nu$ is statistically equivalent, in terms of suitably defined observables, to that of an ensemble of $N$ distinguishable, non-interacting particles, for a time interval $u$ \cite{Breban:2005wf}. A major limitation of the formalism described in this section is that it addresses only the thermodynamics of particle propagation.  This may not be sufficient to account for {\it all} temperature dependence observed in experiment, because the space-time metric may depend on temperature, as well.\footnote{For example, consider the model by which electrons propagate within metal at temperature $T$.  It is natural to assume that the background of electron propagation depends on temperature, which also determines the thermal motion of the electrons.}  This temperature dependence may be included empirically in the metric; here, it remains unaccounted for.  While the role of temperature for the propagation problem defined by Eqs.~\eqref{eq:diracSpin__} and \eqref{eq:Massieu} may not be fully resolved,  our approach allows for addressing other physical aspects, such as the thermodynamic impact of external fields. We discuss two physical systems to clarify the use of our formalism.

\subsection{The free particle} 

For the flat 5D metric, the spin degrees of freedom decouple and may be ignored; Eq.~\eqref{eq:diracSpin__} reduces to an equation for a scalar function $\psi$. A Laplace transform with respect to $y^0$ yields
\begin{eqnarray}
\frac{\partial}{\partial u}\psi'=\frac{c\Lambda}{2}\nabla^2\psi'.
\label{eq:FP.}
\end{eqnarray}
Taking Eq.~\eqref{eq:FP.} for the FP equation relates $\Lambda$ to the diffusion coefficient; i.e., $D=c\Lambda/2$. To link the formalism to thermodynamics, we introduce Kelvin temperature indirectly, through an empirical model of the diffusion coefficient as a function of temperature, $T$. Here, we invoke the Stokes-Einstein model $D=1/(\beta\zeta)$, where $\beta=1/(k_B T)$ and $\zeta$ is the drag coefficient.

The FP equation typically stands for the continuity equation of the probability density to localize a Brownian particle of mass $m$, undergoing non-relativistic dynamics in an environment with temperature $T$ and drag $\zeta$, according to the Langevin equation. We reinterpret the FP equation as the wave equation for a statistical particle of mass $M=\hbar/(2D)$ (i.e., $\Lambda^{-1}\equiv Mc/\hbar$), with the advantage of more flexibility in adding fields to Eq.~\eqref{eq:FP.}.

The wavefunctions $\psi'(x^j,u)=\exp(-e_{\vec{v}}\,u/\hbar) \psi_{\vec{v}}''(x^j)$, where $e_{\vec{v}}=M\vec{v}\,^2/2$ and $\psi_{\vec{v}}''(x^j)=\exp(iM\vec{v}\,\vec{x}/\hbar)$, are eigenstates of $\nabla^2$ in Eq.~\eqref{eq:FP.}. Thus, we may write \cite{Breban:2005wf}
\begin{eqnarray}
\psi'(\vec{x},0; \vec{x},u)=\int d{\vec{v}}\,\psi_{\vec{v}}''(x^j)\exp(-e_{\vec{v}}\,u/\hbar)[\psi_{\vec{v}}''(x^j)]^\dagger=\int  d{\vec{v}}\,\exp(-e_{\vec{v}}\,u/\hbar),
\end{eqnarray}
and
\begin{eqnarray}
\Psi/k_B=\ln \int d{\vec{x}}\,\psi'(\vec{x},0; \vec{x},u).
\end{eqnarray}
We choose $u=2m (\beta D)=2m/\zeta$ \cite{Breban:2005wf}, such that $e_{\vec{v}}\,u/\hbar=\beta m\vec{v}\,^2/2$, and $\Psi/k_B$ yields the traditional thermodynamics for an ideal gas consisting of a single particle.

\subsection{Low-temperature spin in uniform electric and magnetic fields} 

Similarly to Eq.~\eqref{eq:Pauli+spinonly}, we consider weak fields that affect only particle's spin.  We invoke the Arrhenius diffusion model $D\propto \exp(-T_D/T)$ to explicitly request particle diffusion be negligible; i.e., $T\ll T_D$. Still, thermal energy may be stored by the spin degrees of freedom.  Under these conditions, Eq.~\eqref{eq:diracSpin__} reduces to 
\begin{eqnarray}
2\frac{\partial^2\boldsymbol\psi_{\pm}}{\partial y^0\partial y^5}-\frac{q}{c^2}\vec{\boldsymbol\sigma}(i\vec{B}
\mp\vec{E})\left(\frac{\partial}{\partial y^0}+\frac{1}{2}\frac{\partial}{\partial y^5}\right)\boldsymbol\psi_{\pm}\approx 0.
\label{eq:Pauli+spinonly_}
\end{eqnarray}
A Laplace transform with respect to $y^0$ yields
\begin{eqnarray}
-\hbar\frac{\partial\boldsymbol\psi'_{\pm}}{\partial u}\approx-\frac{\mu\vec{\boldsymbol\sigma}(i\vec{B}
\mp\vec{E})\boldsymbol\psi'_{\pm}}{1+(\Lambda/2)q/(2c^2)\vec{\boldsymbol\sigma}(i\vec{B}
\mp\vec{E})}\approx-\mu\vec{\boldsymbol\sigma}(i\vec{B}\mp\vec{E})\boldsymbol\psi'_{\pm}\equiv{\boldsymbol h}\boldsymbol\psi'_{\pm}.
\label{eq:Pauli+spinonly__}
\end{eqnarray}
Hence, we formaly recover Eq.~\eqref{eq:Pauli+spinonly}, where we now require D(1) evolution. The hamiltonian ${\boldsymbol h}=-\mu\vec{\boldsymbol\sigma}(i\vec{B}\mp\vec{E})$ is non-hermitian. The hermitian part of ${\boldsymbol h}$ yields a coherent quantum picture, while the anti-hermitian part of ${\boldsymbol h}$ contributes as a perturbation leading to decoherence. We say that, within the statistical picture, the electric field triggers a quantum coherent response from the particle's spin, while the magnetic field leads to decoherence of the spin state.\footnote{The magnetic field leads to non-hermitian hamiltonians in the statistical pictures of the KG \cite{Breban:2015bf} and Schr\"odinger \cite{Breban:2005wf} equations, as well.}  The converse is valid in the quantum mechanical picture.

The propagator of Eq.~\eqref{eq:Pauli+spinonly__} may be written as
\begin{eqnarray}
\exp(-{\boldsymbol h}u/\hbar)=\exp(-\vec{\boldsymbol\sigma}\vec{\Omega}_{\pm}u)=\cosh(\Omega_{\pm}u)-\frac{\vec{\boldsymbol\sigma}\vec{\Omega}_{\pm}}{\Omega_{\pm}}\sinh(\Omega_{\pm}u),
\end{eqnarray}
where $\vec{\Omega}_{\pm}\equiv \mu(-i\vec{B}\pm\vec{E})/\hbar$ and $\Omega_{\pm}\equiv\sqrt{(\vec{\Omega}_{\pm})^2}\approx\mu(|\vec{E}|\mp i|\vec{B}|\cos\theta)/\hbar$.  Taking the trace of the propagator over the spin and spinor degrees of freedom, we obtain
\begin{eqnarray}
\Psi/k_B=\ln\left[\cosh(\Omega_+u)+\cosh(\Omega_-u)\right]=\ln\left[2\cosh\left(\frac{\mu u}{\hbar}|\vec{E}|\right)\cos\left(\frac{\mu u}{\hbar}|\vec{B}|\cos\theta\right)\right].
\end{eqnarray} 
The electric and magnetic susceptibilities of the spin 1/2 particle are given by
\begin{eqnarray}
\label{eq:chie}
\chi_e&=&\frac{1}{4\pi\beta}\left.\frac{\partial^2(\Psi/k_B)}{\partial |\vec{E}|^2}\right|_{|\vec{E}|=0}=\frac{1}{4\pi\beta}\left(\frac{\mu u}{\hbar}\right)^2,\\
\label{eq:chim}
\chi_m&=&\frac{1}{4\pi\beta}\left.\frac{\partial^2(\Psi/k_B)}{\partial |\vec{B}|^2}\right|_{|\vec{B}|=0}=-\frac{1}{4\pi\beta}\left(\frac{\mu u}{\hbar}\right)^2\cos^2\theta,
\end{eqnarray}
respectively; the case $|\vec{E}|=0$ is formally recovered in Eq.~\eqref{eq:chim} by substituting $\cos^2\theta$ with 1.  Hence, the non-diffusing spin 1/2 particle is paraelectric and diamagnetic, where the time-scale $u$ remains to be established phenomenologically. One possibility is to use Feynman's recipe, $u=\beta\hbar$. However, other choices may be better motivated in applications, and offer a more adapted thermodynamical formalism, particularly that we required $T\ll T_D$. We also required the electric and magnetic fields be sufficiently weak, so their effect on particle orbital motion is negligible. If any of these requirements are violated, the particle may lose its thermodynamic properties, as expressed by Eqs.~\eqref{eq:chie} and \eqref{eq:chim}.

The above results may be important for current discussions on superconductivity.  The reference theory, proposed by Bardeen, Cooper and Schriffer \cite{Bardeen:1957ep,Bardeen:1957hw,Schrieffer:1964tx}, has been challenged for its interpretation of the Meissner effect \cite{Hirsch:2012ga,Hirsch:2017do} and the conduction mechanism \cite{1991PhRvL..67.3448M}. The single-particle thermodynamics developed in this section provide a new starting point to explain the Meissner effect, based on statistical physics with non-hermitian hamiltonians. Furthermore, the superconduction mechanism may rely on interactions between spins of low-temperature particles that are aligned by an external electric field (i.e., {\it electric} spin waves). These remain topics for further work.

\section{Discussion and conclusions}\label{sec:conclusions}

We demonstrated how a 5D formalism of null propagation can provide space-time pictures of spin-1/2, 4D quantum and statistical mechanics.  The resulting 4D hamiltonians in quantum and statistical mechanics may be distinct and non-hermitian. Still, coherent quantum pictures are provided by the hermitian parts of these hamiltonians, where the anti-hermitian parts may be considered as perturbations. Our statistical formalism appears to go against Pauli's theory of paramagnetism.  However, we note that the phenomenology associated with our statistical formalism may be significantly richer than what we presented here.

The minimal requirement for a space-time picture of statistical physics is that the 5D metric be independent of $x^0$ \cite{Breban:2005wf}. Strictly speaking, it is not necessary that the 5D metric be independent of $x^5$ and, implicitly, the mass $m$ of the quantum particle be a well-defined observable.  It is sufficent that the mass $M$, or the concept of diffusion coefficient, be well defined.  Under these circumstances, it is natural to foliate the 5D manifold along $x^0$ and consider the role played by the shift of this foliation in statistical mechanics.  These fields, included in $h_{0M}$, are key for the Lense-Thirring precession and subject of the gravito-electromagnetism theory.  Their microscopic role for statistical phenomena may be carefully considered.

A well-rounded statistical theory may benefit the understanding of other physical phenomena besides superconductivity. Particularly, it may benefit the study of thermodynamic response of condensed matter to external electromagnetic fields. On the quantum mechanical side, the phenomenology of the Dirac equation is well understood. 
Still, experimental confirmation of the anti-hermitian term of spin-electric interaction may provide insight into how to  control the coherence time of quantum devices.


\appendix
\section{Two representations for the $\Gamma$ matrices} \label{Appendix:A}

Sp(1,1) is a pseudo-simplectic group whose generators can be represented in the following form\footnote{We permuted columns 2 and 3, and lines 2 and 3 in the matrix representation of Sp(1,1) generators provided by Ref.~\cite{Helgason:1979vb}.} 
\begin{eqnarray}
\boldsymbol{M}=\left(\begin{array}{cccc}x&b&a&c\\b^*&-x&-c^*&a^*\\-a^*&c&y&d\\-c^*&-a&d^*&-y\end{array}\right),
\label{eq:sp11gen}
\end{eqnarray}
where $x,y\in\mathbb{R}$, $a,b,c,d\in\mathbb{C}$, and $*$ is symbol for complex conjugation.  The Sp(1,1) generators with $a=c=0$ are generators of SU(2)~$\times$~SU(2), subgroup of Sp(1,1), which is a double cover of O(4). It is straightforward to write
four $\Gamma$-matrices that generate the corresponding ${\mathcal Cl_4}$ Clifford subalgebra
\begin{eqnarray}
\boldsymbol\Gamma^j=\left(\begin{array}{cc}\boldsymbol0&i\boldsymbol\sigma^j\\-i\boldsymbol\sigma^j&\boldsymbol0\end{array}\right),\;\boldsymbol\Gamma^5=\left(\begin{array}{cc}\boldsymbol0&-\boldsymbol1\\-\boldsymbol1&\boldsymbol0\end{array}\right),
\label{eq:Gmu}
\end{eqnarray}
where $\boldsymbol\sigma^j$ are the Pauli matrices ($j,k, ...=1,2,3$) and $i\boldsymbol\sigma^j$ is a representation of the quaternion units. The $\Gamma$ matrices may be considered $2\times 2$ with complex quaternion entries or $4\times 4$ with complex entries. They satisfy $\Gamma^{\alpha a}{}_b\Gamma^{\beta b}{}_c+\Gamma^{\beta a}{}_b\Gamma^{\alpha b}{}_c=2(1^{\alpha\beta}1^a{}_c)$, where the indices $\alpha,\beta=1,2,3,5$ and $a,b,c=1,2$ or $1,2,3,4$, as needed. The generators of SU(2)~$\times$~SU(2) can be represented using commutators of $\Gamma$ matrices 
\begin{eqnarray}
\boldsymbol\sigma^{jk}\equiv i[\boldsymbol\Gamma^j,\boldsymbol\Gamma^k]/2&=&-\epsilon^{jk}{}_l\left(\begin{array}{cc}\boldsymbol\sigma^l&\boldsymbol0\\\boldsymbol0&\boldsymbol\sigma^l\end{array}\right),\\
\boldsymbol\sigma^{j5}\equiv i[\boldsymbol\Gamma^j,\boldsymbol\Gamma^5]/2&=&\left(\begin{array}{cc}\boldsymbol\sigma^j&\boldsymbol0\\\boldsymbol0&-\boldsymbol\sigma^j\end{array}\right).
\end{eqnarray}
We now introduce
\begin{eqnarray}
\boldsymbol\Gamma^0\equiv i\boldsymbol\Gamma^1\boldsymbol\Gamma^2\boldsymbol\Gamma^3\boldsymbol\Gamma^5=i\left(\begin{array}{cc}\boldsymbol1&\boldsymbol0\\\boldsymbol0&-\boldsymbol1\end{array}\right).
\label{eq:G0}
\end{eqnarray}
Direct calculations provide $\Gamma^{Aa}{}_b\Gamma^{Bb}{}_c+\Gamma^{Ba}{}_b\Gamma^{Ab}{}_c=2\eta^{AB}1^a{}_c$, and the following commutation relations
\begin{eqnarray}
\boldsymbol\sigma^{0j}\equiv i[\boldsymbol\Gamma^0,\boldsymbol\Gamma^j]/2&=&\left(\begin{array}{cc}\boldsymbol0&-i\boldsymbol\sigma^j\\-i\boldsymbol\sigma^j&\boldsymbol0\end{array}\right),\\
\boldsymbol\sigma^{05}\equiv i[\boldsymbol\Gamma^0,\boldsymbol\Gamma^5]/2&=&\left(\begin{array}{cc}\boldsymbol0&\boldsymbol1\\-\boldsymbol1&\boldsymbol0\end{array}\right),
\end{eqnarray}
which are the other four Sp(1,1) generators, obtained by taking $b=d=x=y=0$ in \eqref{eq:sp11gen}. The $\Gamma$ matrices given by Eqs.~\eqref{eq:Gmu} and \eqref{eq:G0} satisfy 
\begin{eqnarray}
\mathrm{Tr}\,(\boldsymbol\Gamma_A)=0,
\end{eqnarray}
and have the desired properties under hermitian conjugation  
\begin{eqnarray}
\boldsymbol\Gamma_0^\dagger=-\boldsymbol\Gamma_0, \quad \boldsymbol\Gamma_j^\dagger=\boldsymbol\Gamma_j, \quad \boldsymbol\Gamma_5^\dagger=\boldsymbol\Gamma_5.  
\label{eq:hermitian2}
\end{eqnarray}
Thus, Eqs.~\eqref{eq:Gmu} and \eqref{eq:G0} provide a representation of the basis of ${\mathcal Cl^s}_{1,4}$. Finally, defining $\boldsymbol\Gamma\equiv ip\boldsymbol\Gamma^0+q\boldsymbol\Gamma^1\boldsymbol\Gamma^2\boldsymbol\Gamma^3\boldsymbol\Gamma^5=-ip\boldsymbol\Gamma_0+q\boldsymbol\Gamma_1\boldsymbol\Gamma_2\boldsymbol\Gamma_3\boldsymbol\Gamma_5$, where $p,q\in\mathbb{R}$ (n.b., $\boldsymbol\Gamma^\dagger=\boldsymbol\Gamma$), we obtain 
\begin{eqnarray}
\boldsymbol\Gamma\boldsymbol\Gamma^\dagger_A=-\boldsymbol\Gamma_A\boldsymbol\Gamma
\label{eq:Gamma_}
\end{eqnarray}
and 
\begin{eqnarray}
(\boldsymbol\Gamma)^2=(p-q)^2\boldsymbol 1.
\label{eq:Gamma__}
\end{eqnarray}

For applications, we note that the matrices $\boldsymbol\gamma^\alpha\equiv\boldsymbol\Gamma^5\boldsymbol\Gamma^{\alpha}$, given by 
\begin{eqnarray}
\boldsymbol\gamma^0=\left(\begin{array}{cc}\boldsymbol0&i\boldsymbol 1\\-i\boldsymbol 1&\boldsymbol0\end{array}\right),\;\boldsymbol\gamma^j=\left(\begin{array}{cc}i\boldsymbol\sigma^j&\boldsymbol0\\\boldsymbol0&-i\boldsymbol\sigma^j\end{array}\right),
\end{eqnarray}
form the basis for a 4D Clifford algebra where $\gamma^{\alpha a}{}_b\gamma^{\beta b}{}_c+\gamma^{\beta a}{}_b\gamma^{\alpha b}{}_c=-2\eta^{\alpha\beta}1^a{}_c$ and $\alpha,\beta=0,1,2,3$. Furthermore, for covariance breaking, we may use a different representation of the $\Gamma$ matrices, generated by conjugacy $\boldsymbol\Gamma'^A=\boldsymbol U\Gamma^A\boldsymbol U^{-1}$ with the unitary matrix  
\begin{eqnarray}
\boldsymbol U=\frac{1}{\sqrt{2}}\left(\begin{array}{cc}-\boldsymbol1&i\boldsymbol 1\\-i\boldsymbol 1&\boldsymbol1\end{array}\right).
\end{eqnarray}
We obtain 
\begin{eqnarray}
\boldsymbol\Gamma'^0=\left(\begin{array}{cc}\boldsymbol0&\boldsymbol 1\\-\boldsymbol 1&\boldsymbol0\end{array}\right),\;
\boldsymbol\Gamma'^j=\left(\begin{array}{cc}-\boldsymbol\sigma^j&\boldsymbol 0\\\boldsymbol 0&\boldsymbol\sigma^j\end{array}\right),\;
\boldsymbol\Gamma'^5=\left(\begin{array}{cc}\boldsymbol0&\boldsymbol 1\\ \boldsymbol 1&\boldsymbol0\end{array}\right).
\end{eqnarray}
The matrices $\boldsymbol\gamma'^\alpha\equiv\boldsymbol\Gamma'^5\boldsymbol\Gamma'^{\alpha}$ are given by
\begin{eqnarray}
\boldsymbol\gamma'^0=\left(\begin{array}{cc}-\boldsymbol1&\boldsymbol 0\\\boldsymbol 0&\boldsymbol1\end{array}\right),\;\boldsymbol\gamma'^j=\left(\begin{array}{cc}\boldsymbol0&\boldsymbol\sigma^j\\-\boldsymbol\sigma^j&\boldsymbol0\end{array}\right).
\label{eq:gamma'}
\end{eqnarray}
In the main text, we refer to this representation as the primed representation of the gamma matrices.  The Dirac $\gamma$-matrices are obtained from Eq.~\eqref{eq:gamma'} by changing the sign of $\boldsymbol\gamma'^0$; i.e., $\boldsymbol\gamma'^0\rightarrow -\boldsymbol\gamma'^0$.

\section{The biquaternion-bispinor relationship}\label{Appendix:B}

Given a bispinor $\boldsymbol\psi_1$, there exists a unique biquaternion $\boldsymbol\Psi$ which, represented with the hypercomplex units $i\boldsymbol\sigma^j$, has the bispinor $\boldsymbol\psi_1$ as its first column; i.e., $\boldsymbol\Psi=(\boldsymbol\psi_1~\boldsymbol\psi_2)$.  Straightforward calculations yield $\boldsymbol\psi_2=\boldsymbol K\boldsymbol\psi_1^*$, where $*$ is symbol for complex conjugation, and $\boldsymbol{K}$ is a $4\times 4$ antisymmetric matrix with real entries
\begin{eqnarray}
\boldsymbol{K}=\left(\begin{array}{cccc}0&-1&0&0\\1&0&0&0\\0&0&0&-1\\0&0&1&0\end{array}\right)=-i\left(\begin{array}{cc}\boldsymbol\sigma^2&\boldsymbol0\\\boldsymbol0&\boldsymbol\sigma^2\end{array}\right).
\end{eqnarray}
A biquaternion of unit norm
\begin{eqnarray}
\boldsymbol\Psi\cdot\boldsymbol\Psi=\left(\begin{array}{c}\boldsymbol\psi^\dagger_1\\\boldsymbol\psi^\dagger_2\end{array}\right)\boldsymbol\Gamma\left(\boldsymbol\psi_1~\boldsymbol\psi_2\right)=\left(\begin{array}{cc}\boldsymbol\psi_1\cdot\boldsymbol\psi_1 & \boldsymbol\psi_1\cdot\boldsymbol\psi_2\\ \boldsymbol\psi_2\cdot\boldsymbol\psi_1 & \boldsymbol\psi_2\cdot\boldsymbol\psi_2\end{array}\right)=\left(\begin{array}{cc}\boldsymbol1&\boldsymbol0\\\boldsymbol0&\boldsymbol1\end{array}\right),
\end{eqnarray}
 is equivalent to two mutually orthonormal bispinors $\boldsymbol\psi_1$ and $\boldsymbol\psi_2$; n.b., $\boldsymbol\psi_1\cdot\boldsymbol\psi_2=0$ is identically satisfied, owing to the relationship between $\boldsymbol\psi_1$ and $\boldsymbol\psi_2$.

A quaternion-spinor relationship can be established in similar fashion. For each spinor $\boldsymbol\psi_{11}$, there exists a unique quaternion $\boldsymbol\Psi_1$ which, represented with the hypercomplex units $i\boldsymbol\sigma^j$, has the spinor $\boldsymbol\psi_{11}$ as its first column; i.e., $\boldsymbol\Psi_1=(\boldsymbol\psi_{11}~\boldsymbol\psi_{12})$, where $\boldsymbol\psi_{12}=-i\boldsymbol\sigma^2\boldsymbol\psi_{11}^*$ so $\boldsymbol\psi_{11}^\dagger\boldsymbol\psi_{12}=0$.  If $\boldsymbol\Psi_1$ has unit norm, we have 
\begin{eqnarray}
\boldsymbol\Psi_1\cdot\boldsymbol\Psi_1=\boldsymbol\Psi_1^\dagger\boldsymbol\Psi_1=\left(\begin{array}{c}\boldsymbol\psi^\dagger_{11}\\\boldsymbol\psi^\dagger_{12}\end{array}\right)\left(\boldsymbol\psi_{11}~\boldsymbol\psi_{12}\right)=\left(\begin{array}{cc}1&0\\0&1\end{array}\right),
\end{eqnarray}
implying that $\boldsymbol\psi_{11}$ and $\boldsymbol\psi_{12}$ are mutually orthogonal and have unit norm, as well.

\end{document}